\documentclass[12pt]{article}
\usepackage{graphicx}
\usepackage{wrapfig}

\begin{document}

\title{$\eta$ MAID-2015: update with new data and new resonances}


\author{V.L. Kashevarov, L. Tiator, M. Ostrick\\
Institut f\"ur Kernphysik, Johannes Gutenberg-Universit\"at\\
D-55099 Mainz, Germany}

\maketitle

\begin{abstract}

Recent data for $\eta$ and $\eta\prime$ photoproduction on protons 
obtained by the A2 Collaboration at MAMI are presented. 
The total cross section for $\eta$ photoproduction demonstrates a cusp at the energy 
corresponding to the $\eta\prime$ threshold. 
The new data and existing data from GRAAL, CBELSA/TAPS, and CLAS collaborations have 
been analyzed by an expansion in terms of associated Legendre polynomials.
The isobar model $\eta$MAID updated with $\eta \prime$ channel and new resonances have
been used to fit the new data. The new solution $\eta$MAID-2015 reasonably good  
describes the data in the photon beam energy region up to 3.7 GeV. 

\end{abstract}

\section{Introduction}

The unitarity isobar model $\eta$MAID \cite{MAID} was developed in 2002 for $\eta$ photo- 
and electroproduction on nucleons. The model includes a nonresonant background, which consists of
nucleon Born terms in the $s$ and $u$ channels and the vector meson exchange in the $t$ channel,
and $s$-channel resonance excitations. The Born terms are evaluated with the pseudoscalar coupling.
The vector meson contribution is obtained by the $\rho$ and $\omega$ meson exchange in 
the $t$ channel with pole-like Feynman propagators. For each 
partial wave the resonance contribution is parameterized by the Breit-Wigner function
with energy dependent widths. The $\eta$MAID-2003 version includes 8 resonances, $N(1520)3/2^-$,
$N(1535)1/2^-$, $N(1650)1/2^-$, $N(1675)5/2^-$, $N(1680)5/2^+$, $N(1700)3/2^-$, $N(1710)1/2^+$, 
$N(1720)3/2^+$,
and was fitted to proton data for differential cross sections and beam asymmetry at photon 
beam energies up to 1400 MeV. The $\eta$MAID-2003 version describes not only 
the experimental data available in 2002, but even a bump structure around W=1700 MeV in
$\eta$ photoproduction on the neutron, which was observed a few years later.     
However, this version fails to reproduce the new polarization data obtained in Mainz
\cite{TF_MAMI}.    

The aim of this work is to extend the $\eta$MAID-2003 version to higher energies, to improve
a description of the new polarization data, and to include the $\eta \prime$ photoproduction 
channel. 

\section{Truncated Legendre analysis}

The full angular coverage of differential cross sections and polarization observables
allow us to perform a fit with a Legendre series truncated to a maximum orbital angular
momentum $\ell_{\mathrm{max}}$:

\begin{eqnarray}
\frac{d\sigma}{d\Omega} &=&
\sum\limits_{n=0}^{2 \ell_{\mathrm{max}}} A^{\sigma}_{n}P^0_{n}(\cos\Theta_{\eta}),  \\\label{LegPol2}
T (F) \;\; \frac{d\sigma}{d\Omega} &=&
\sum\limits_{n=1}^{2 \ell_{\mathrm{max}}} A^{T(F)}_{n}P^1_{n}(\cos\Theta_{\eta}), \\\label{LegPol1}
\Sigma \;\; \frac{d\sigma}{d\Omega} &=&
\sum\limits_{n=2}^{2 \ell_{\mathrm{max}}} A^{\Sigma}_{n}P^2_{n}(\cos\Theta_{\eta}),  \label{LegPol4}
\end{eqnarray}
where $P^m_{n}(\cos\Theta_{\eta})$ are associated Legendre polynomials.
The spin-dependent cross sections, $T d\sigma/d\Omega$, $F d\sigma/d\Omega$,
and $\Sigma d\sigma/d\Omega$ were obtained by multiplying the corresponding asymmetries with the
differential cross sections obtained in Mainz. 
\begin{figure*}
\begin{center}
\resizebox{0.8\textwidth}{!}{\includegraphics{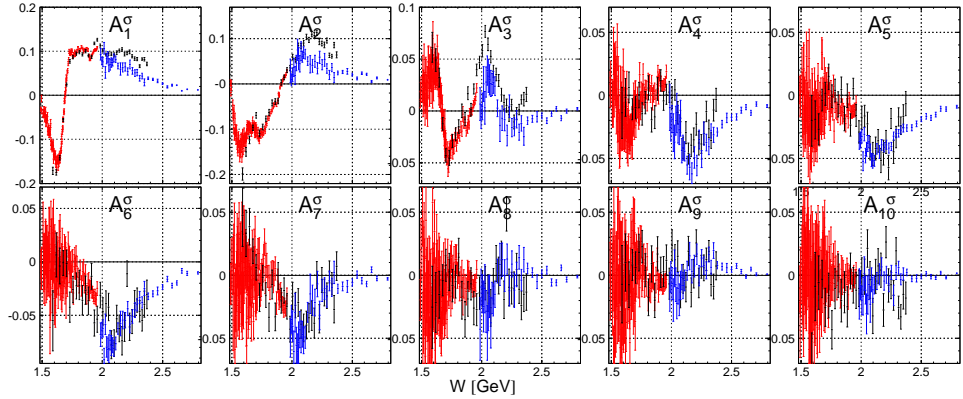}}
\caption{Legendre coefficients in $[\mu b/sr]$ up to $\ell_{\mathrm{max}} = 5$ from
our fits to the differential cross section of the $\gamma p \to \eta p$ reaction
as function of the center-of-mass energy W. 
Red circles are fit results for preliminary A2MAMI data \cite{dcs_MAMI}, black and 
blue - for CBELSA/TAPS \cite{dcs_ELSA} and CLAS \cite{dcs_CLAS} data correspondingly.
}
\label{fig1}
\end{center}
\end{figure*}
\begin{figure*}
\begin{center}
\resizebox{0.8\textwidth}{!}{\includegraphics{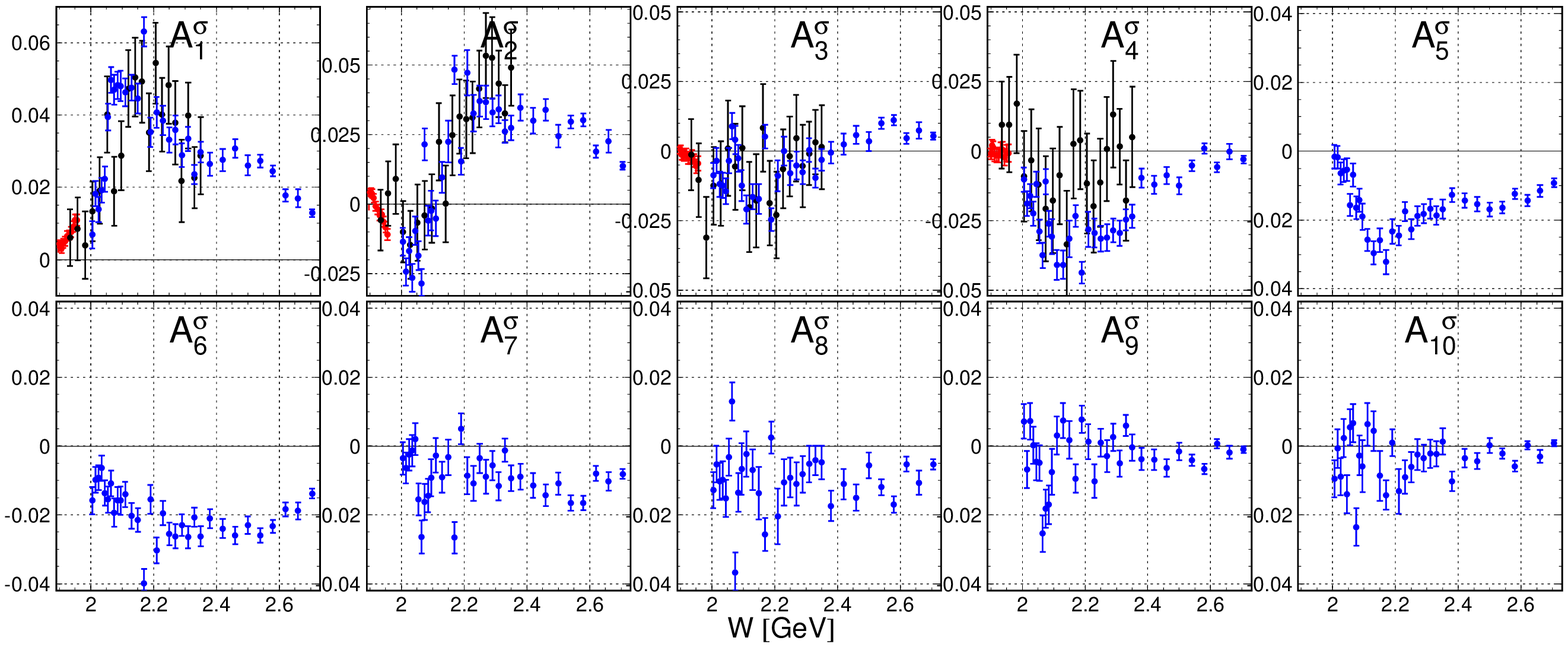}}
\caption{The same as Fig.\,\ref{fig1}, but for the $\gamma p \to \eta \prime p$ reaction.
}
\label{fig2}
\end{center}
\end{figure*}
As an example, the results for the Legendre coefficients for differential cross sections 
are presented in Figs.\,\ref{fig1} and \ref{fig2}.
A non-zero $A_{10}$ only posible with $h$-wave contribution,
$A_9$ is dominated by the an interference between $g$ and $h$ waves, $A_8$ includes $g$, $h$ waves and an interference
between $f$ and $h$ waves, and so on.
The first coefficient, $A_0$, was omitted in the figures because of it includes all 
possible partial-wave amplitudes and just only reflects the magnitude of the total cross section, see Fig.~3. 

Non-zero values of the $A_7$ and $A_8$ coefficients point to a contribution of the $g$ wave at energies
above W=2 GeV for both $\eta$ and $\eta \prime$ channels.   
The errors in the determination of the coefficients $A_9$ and $A_{10}$ do not allow any conclusions
about the contribution of $h$ wave in these reactions. Polarization observables for $\eta$ photoprduction
were measured below W=1.9 GeV. The Legendre fit for these data shows the sensitivity to small partial-wave
contributions and indicates $pd$ interferences below W=1.6 GeV and $df$ interferences above 
W=1.6 GeV \cite{TF_MAMI}. 
 
\section{Updated $\eta$MAID}

New $\eta$MAID-2015 model is based on the $\eta$MAID-2003 version. 
The following main changes were made: 
\begin{itemize}
\item 12 additional resonances were added: $N(1860)5/2^+$,
$N(1875)3/2^-$, $N(1880)1/2^+$, $N(1895)1/2^-$, $N(1900)3/2^+$, $N(1990)7/2^+$, $N(2000)5/2^+$,
$N(2060)5/2^-$, $N(2120)3/2^-$, $N(2190)7/2^-$, $N(2220)9/2^+$, and $N(2250)9/2^-$;
\item electromagnetic couplings for the vector mesons were updated according to Ref.~\cite{PDG-14};
\item hadronic vector and tensor couplings for the vector mesons were fixed from Ref.~\cite{Laget-05};
\item data base for the fit was updated. 
\end{itemize}

The new model was fitted to data of differential cross sections from A2MAMI \cite{dcs_MAMI} 
and CLAS Collaborations \cite{dcs_CLAS}, polarisation observables T, F \cite{TF_MAMI} and 
$\Sigma$ \cite{GRAAL-07}, \cite{GRAAL-15}. 
The main variable parameters for each resonance: Breit-Wigner mass, 
total width, branching ratio to $\eta p$ (or $\eta \prime p$) decay, photoexcitation helicity 
amplitudes $A_{1/2}$ and $A_{3/2}$, a relative sign between the $N^* \to \eta N$ and the 
$N^* \to \pi N$ couplings. Besides, the hadronic pseudoscalar coupling for the Born term contribution,
cutoffs for dipole formfactors of the vector mesons, damping factors for the partial widths 
and the electromagnetic form factor of the resonances were also fitted. 
Branching ratios for hadronic decays of the resonances besides the investigated channel were fixed. 

As an initial parameter set for the Breit-Wigner parameters the last BnGa solution \cite{BnGa} was used.
As initial parameter limits uncertainties from Refs.~\cite{PDG-14} and \cite{BnGa} were used.
As the first step, for each resonance $A_{1/2}$ and $A_{3/2}$ are fixed because of a strong correlation with the branching
ratio. On the second step the branching ratios obtained on the first step are fixed, but $A_{1/2}$ and $A_{3/2}$
are variable, and so on. After few iterations the initial limits are changed if necessary. The fits for
the $\eta$ and $\eta \prime$ channels were done independently.
         
The fit results for the total cross sections and the polarization observables are presented 
in Figs.~3-6 together with corresponding experimental data.
We used the differential cross section from the CLAS Collabration~\cite{dcs_CLAS} in this fit because of their
much smaller statistical errors, larger energy covering, 
and better agreement with the high statistic data from A2MAMI~\cite{dcs_MAMI} in an overlapping energy region.
Unfortunately, the total cross section was not determined in Ref.~\cite{dcs_CLAS} and we calculated it
using Legendre decomposition for the differential cross sections. Blue circles in Figs.~3 and 5 are
results of this procedure.

In Fig.~3, there is a very interesting feature at energy $\sim$1900 MeV, which could be explained by a 
cusp due to the opening of a new channel, $\eta \prime$ photoproduction.
The main resonance, which is responsible for this effect is the $N(1895)1/2^-$. The Breit-Wigner parameters 
of this state were determined by the fit as following: 
$M=1896 \pm 1$ MeV, $\Gamma_{tot}=93 \pm 13$ MeV, $\Gamma_{\eta p}=(14 \pm 3) \%$,
$\Gamma_{\eta\prime p}=(6.5 \pm 2) \%$, and $A_{1/2}=(-17.4 \pm 1.5) 10^{-3} GeV^{-1/2}$.  
Fig.~4 demonstrates a significant improvement of description for T and F asymmetries (red lines) in comparison
with the $\eta$MAID-2003 version (blue lines).
\begin{figure}
\begin{center}
\includegraphics[width=0.9\textwidth]{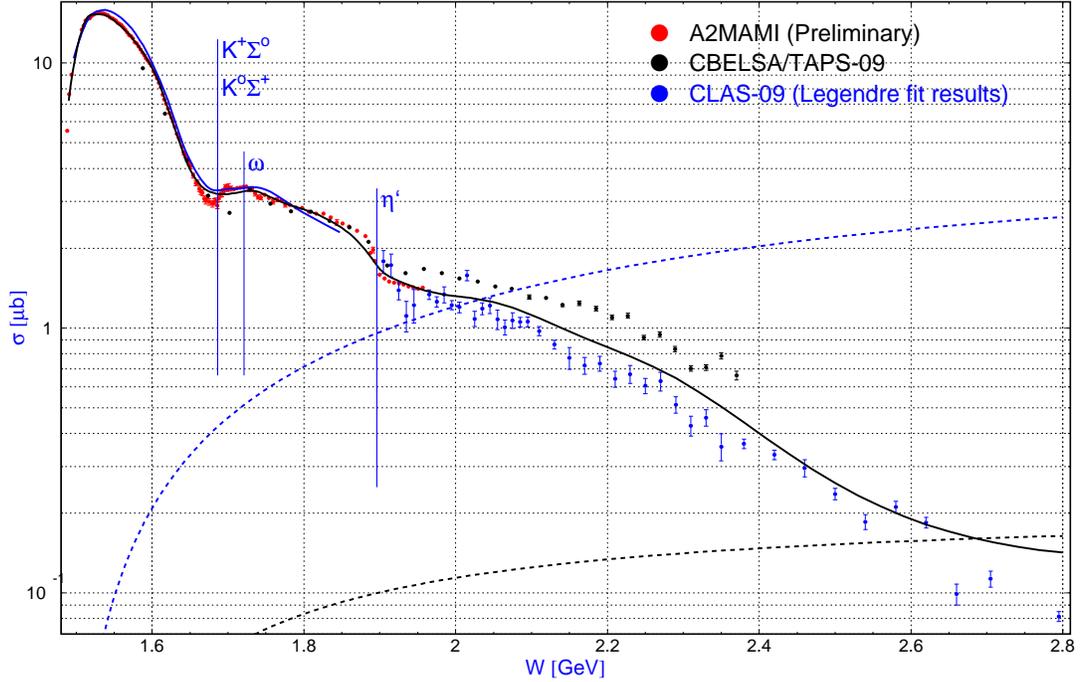}
\caption{Total cross section of the $\gamma p \to \eta p$ reaction.
Solid blue curve is $\eta$MAID-2003 isobar model \cite{MAID},
black solid curve: new $\eta$MAID-2015 solution.
Prediction of $\eta$MAID-2003 for background contribution is shown by blue dashed line,
background of $\eta$MAID-2015 - black dashed line.   
Vertical lines correspond to thresholds of $K\Sigma$, $\omega$, and $\eta \prime$ 
photoproductions.
}
\label{fig3}
\end{center}
\end{figure}
\begin{figure}
\begin{center}
\includegraphics[width=0.8\textwidth]{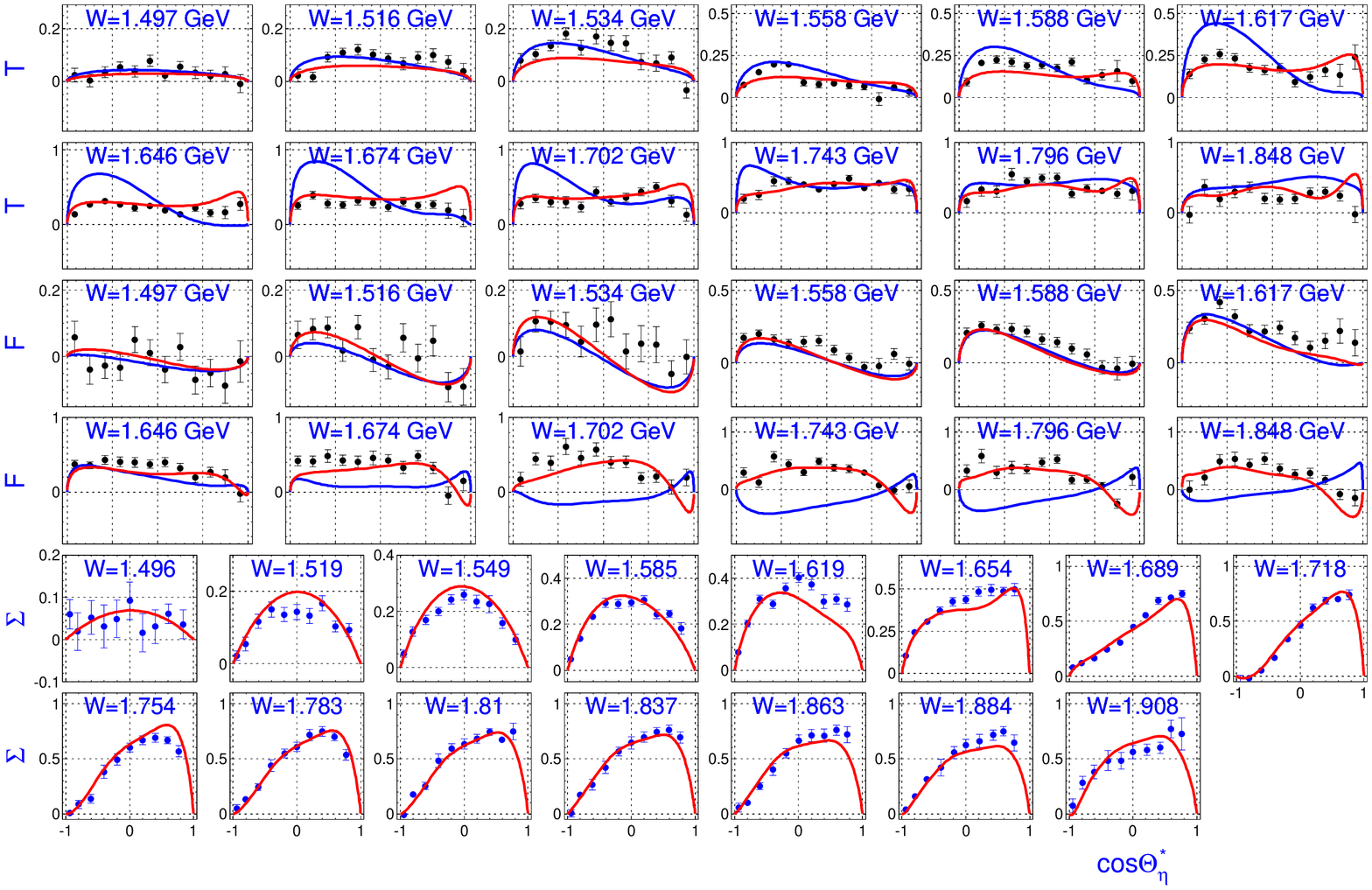}
\caption{$\eta$MAID-2015 solution for the $\eta$ channel (red lines). 
Black circles: A2MAMI-15 data \cite{TF_MAMI} for T and F asymmetries, 
blue circles: GRAAL-07 data \cite{GRAAL-07} for $\Sigma$.
Blue lines: $\eta$MAID-2003 prediction \cite{MAID}.
}
\label{fig4}
\end{center}
\end{figure}

A very good agreement with the experimental data was obtained for the cross section of the 
$\gamma p \to \eta \prime p$ reaction (see Fig.~5). The main contributions to this reaction come from
$N(1895)1/2^-$, $N(1900)3/2^+$, $N(1880)1/2^+$, $N(2150)3/2^-$, and $N(2000)5/2^+$ resonances.
Other resonance contributions are much smaller then the background.  
The new $\eta$MAID-2015 solution describes shape of the GRAAL data for $\Sigma$ near threshold,
but not the magnitude (see Fig.~6). To explain, why the magnitude of the asymmetry is larger at lower
energy, it is probably necessary to include below threshold resonances using the more realistic  
approach applied in Ref.~\cite{Azn-03} for the Roper resonance at $\eta$-meson photoproduction. 
 
\begin{figure}
\begin{center}
\resizebox{0.7\textwidth}{!}{\includegraphics{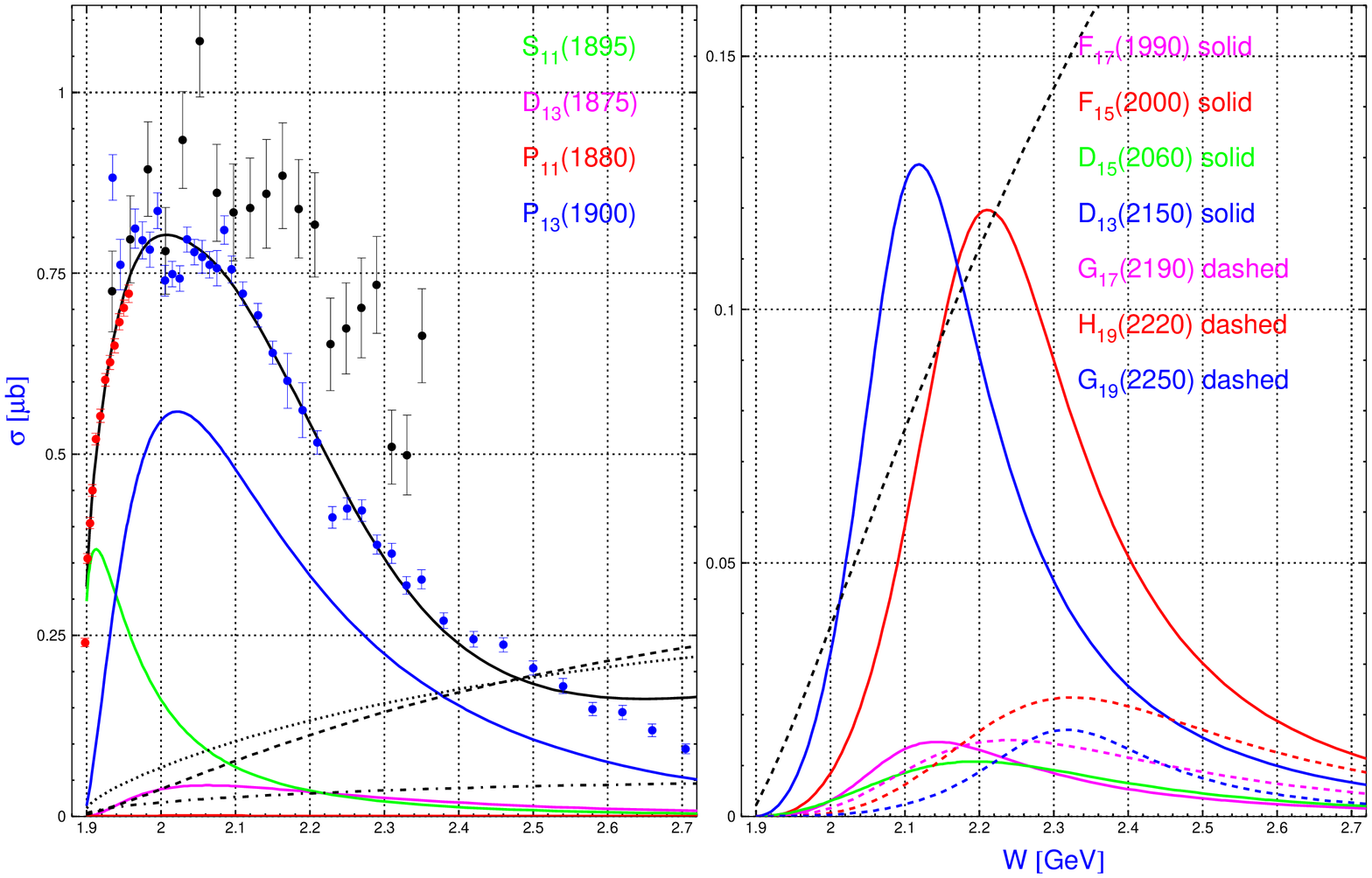}}
\caption{Total cross section of the $\gamma p \to \eta\prime p$ reaction. 
Red circles: A2MAMI-15 data \cite{dcs_MAMI}, black circles: CBELSA/TAPS-09 
\cite{dcs_ELSA}, blue circles: data obtained from the Legendre fit to the 
differential cross sections of the CLAS Collaboration \cite{dcs_CLAS}. 
Solid black line: $\eta$MAID-2015 solution. Background contribution is
shown by dashed black line. Black dotted and dot-dashed lines are partial
contributions of the Born terms and the vector mesons correspondingly. 
Other curves are partial contributions of resonances.  
}
\label{fig5}
\end{center}
\end{figure}
\begin{figure}
\begin{center}
\resizebox{0.25\textwidth}{!}{\includegraphics{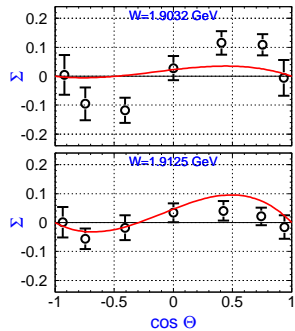}}
\caption{Beam asymmetry $\Sigma$. Data from Ref.\cite{GRAAL-15}, red curves are
$\eta$MAID-2015 solution.}
\label{fig6}
\end{center}
\end{figure}

\section{Summary and conclusions}

In summary, we have presented new version $\eta$MAID-2015.
The model describes available data for the $\gamma p \to \eta p$ and 
$\gamma p \to \eta \prime p$ reactions reasonably well.
The cusp at W$\sim$1900 MeV in $\gamma p \to \eta p$ reaction was explained as a threshold effect
from the $\eta \prime$ channel. Parameters of $N(1895)1/2^-$ resonance, 
responsible for this effect, were determined.    
A further improvement could be achieved by adding below threshold resonances
and using Regge trajectories for the vector mesons in $t$ channel.
Furthermore, polarization observables which should come soon 
from A2MAMI, CBELSA/TAPS, and CLAS Collaborations will help to improve the model.

This work was supported by the Deutsche Forschungsgemeinschaft (SFB 1044).


\end{document}